\input{aipcheck}
\documentclass[
    ,final            
  ]
  {aipproc}

\layoutstyle{6x9}

\begin{document}

\title[Time Variability of Quasars]{Time Variability of Quasars: the Structure Function Variance}


\author{C.~MacLeod}{
  address={University of Washington, Seattle, USA}
}
\author{\v{Z}. Ivezi\'{c}}{
  address={University of Washington, Seattle, USA}
}
\author{W. de Vries}{
  address={University of California, Davis, USA}
}
\author{B. Sesar}{
  address={University of Washington, Seattle, USA}
}
\author{A. Becker}{
  address={University of Washington, Seattle, USA}
}

\begin{abstract}
Significant progress in the description of quasar variability has 
been recently made by employing SDSS and POSS data. Common to
most studies is a fundamental assumption that photometric
observations at two epochs for a large number of quasars will
reveal the same statistical properties as well-sampled 
light curves for individual objects. We critically test this
assumption using light curves for a sample of $\sim$2,600
spectroscopically confirmed quasars observed about 50 times on 
average over 8 years by the SDSS stripe 82 survey. 
We find that the dependence of the mean structure function computed
for individual quasars on luminosity, rest-frame wavelength and 
time is qualitatively and quantitatively similar to the behavior 
of the structure function derived from two-epoch observations of 
a much larger sample. We also reproduce the result that the
variability properties of radio and X-ray selected subsamples
are different. However, the scatter of the variability structure function 
for fixed values of luminosity, rest-frame wavelength and time is 
similar to the scatter induced by the variance of these quantities
in the analyzed sample. Hence, our results suggest that, although the 
statistical properties of quasar variability inferred using two-epoch
data capture some underlying physics, there is significant additional 
information that can be extracted from well-sampled light curves for 
individual objects. 
\end{abstract}

\maketitle


\section{Introduction}

Significant progress in the description of quasar variability has been recently made by 
employing SDSS data. Vanden Berk et~al.\ (\cite{VB2004}, hereafter VB) compared 
imaging and spectrophotometric magnitudes to investigate the correlations of 
variability with rest-frame time lag (up to 2 years), luminosity, rest-frame wavelength, 
redshift, the presence of radio and X-ray emission, and the presence of broad absorption 
line outflows. Variability on longer time scales was studied by de
Vries, Becker, White, and Loomis 
(\cite{dV2005}, hereafter dVBWL; \cite{Sesar2006}) who compared SDSS and POSS photometric 
measurements. Ivezi\'{c} et~al.\ (\cite{ive2004}, hereafter, I04) used repeated SDSS 
imaging scans, which increased the measurement accuracy for magnitude differences by 
about a factor of 3-4 compared to
studies based on spectrophotometric magnitudes, and also enabled analysis of the 
$u$ and $z$ band variability. 

Common to these and similar studies is a fundamental assumption that photometric
observations at two epochs for a large number of quasars will reveal the same statistical 
properties as well-sampled light curves for individual objects. We critically test this
assumption using light curves for a sample of $\sim$2,600 spectroscopically confirmed quasars 
observed by the multi-epoch SDSS stripe 82 survey. We describe these observations
and our methodology in the next section, and analyze the data in Section 3.

\section{SDSS Stripe 82 Data and Methodology}

We study a sample of about 2,600 spectroscopically confirmed quasars
(see \cite{Sch2007} for the SDSS Quasar Catalog) imaged about 50 times on the average
over 8 years (for details
see \cite{Sesar2007}). These data were obtained in yearly ``seasons'' about
2-3 months long. We average observations within each season, and construct 
a structure function (see dVBWL for definition) for each object for an observed 
time lag of 1 year, and in each of the five SDSS bandpasses ($ugriz$). The
total number of structure function data points for all quasars and bands is 
10,370. We correct for cosmological time dilation and length contraction and 
obtain a fairly well-sampled plane of the rest-frame quantities $\Delta
t_{RF}$ and $\lambda_{RF}$ (spanning from about 
100 to 300 days and 1000 to 6000 \AA).  

\begin{figure}
\includegraphics[width=6in,viewport=0 132 400 259,clip]{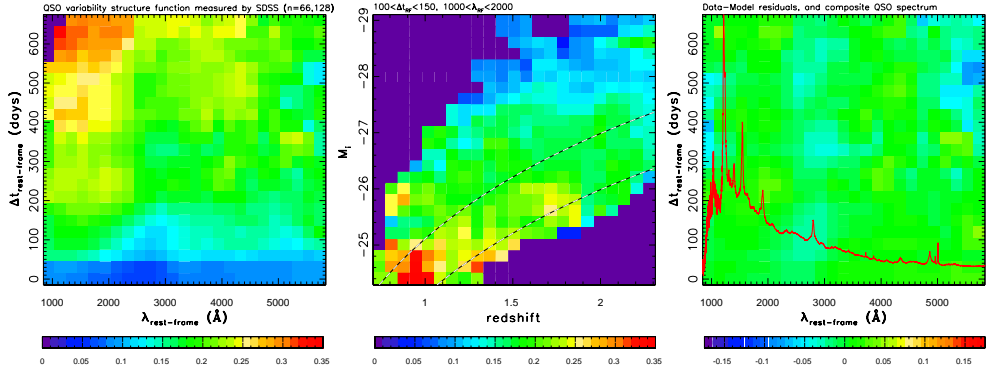}
\caption{(Figure 1 from I04) The left panel displays the measured structure function as a function of rest-frame 
time lag and wavelength (each pixel contains $\sim25$ objects). The middle panel
shows the structure function in a narrow range of rest-frame time lag and wavelength
(100 days$<\Delta t_{RF}<$150 days, 1000\AA $<\lambda_{RF}<$2000 \AA),
as a function of redshift and luminosity. The lines of constant variability are nearly 
parallel to the redshift axis, suggesting that the dependence of variability
on luminosity is much stronger than the dependence on redshift. The right panel
displays the difference between the data shown in the left panel and the best-fit
model described by Eq.~(\ref{eq:eq1}). The over-plotted line shows the composite quasar spectrum 
from \cite{VB2001}. The negative residuals at $\sim$2800
\AA\ are due to the anti-correlation of continuum and MgII line variability.}
\label{fig:2epoch}
\end{figure}

To compare the statistical properties of quasar variability inferred from 
two-epoch data, and those based on light curves for individual objects,
we use a quasar variability model from I04. Using $\sim$66,000 magnitude 
difference measurements, $\Delta m$, for a sample of $\sim$13,000 quasars, 
they found that $\Delta m$ follow an exponential distribution, 
$p(\Delta m)\propto exp(-|\Delta m|/\Delta_c)$, where the characteristic
variability scale, $\Delta_c$ is a function of rest-frame time lag 
($\Delta t_{RF}$, days), wavelength ($\lambda_{RF}$, \AA), and absolute magnitude 
in the $i$ band ($M_i$) (the variability scale is related to the more commonly used 
structure function by $SF=\sqrt{2}\Delta_c$). Their simple model for the variability
scale,

\begin{equation}
SF = (1.00\pm0.03)[1 + (0.024\pm0.04)\,M_i]  \left(\frac{\Delta t_{RF}}{\lambda_{RF}}\right)^{0.30\pm0.05}
\label{eq:eq1}
\end{equation}
describes $\Delta m$ measurements to within the measurement noise
($\sim$0.02 mag). Note that
there is no dependence on redshift (see Fig.~\ref{fig:2epoch}).

We attempted to fit the same functional form to the SF data obtained for
individual objects from stripe 82. We find that the observed SF is
still well-fit using a single exponent for $\lambda_{RF}$ and $\Delta
t_{RF}$, except with a value for the exponent of $0.47\pm0.02$. The
parameters used to model the observed SF are measured in the ranges
1000\AA~< $\lambda_{RF}$<~6000\AA, 100 days $< \Delta
t_{RF} < 300$ days (compared to 0-600 days for the two-epoch
sample), and $-29 < M_i < -23$. This
modified model SF provides a good description of the observed SF; the
ratio of the observed to model SF does not show any systematic
behavior with respect to $\Delta t_{RF}$, $\lambda_{RF}$, $M_i$, or
redshift.  
 
The similarity of the {\it mean} structure function computed for individual quasars 
and the structure function derived from the two-epoch observations of 
a much larger sample suggests that the statistical properties 
of quasar variability inferred using only two epochs for a large 
number of quasars are representative of the underlying physics. This
is at least true for the overlap in their timescales (i.e., for
timescales $\sim$300 days or less). We 
proceed with analysis of the distribution of the ratio of observed ($SF_{obs}$) 
and modeled ($SF_{model}$) structure functions.

\section{       Analysis: the Structure Function Variance  }

\begin{figure}[htbp]
   \centering
\includegraphics[width=5.3in]{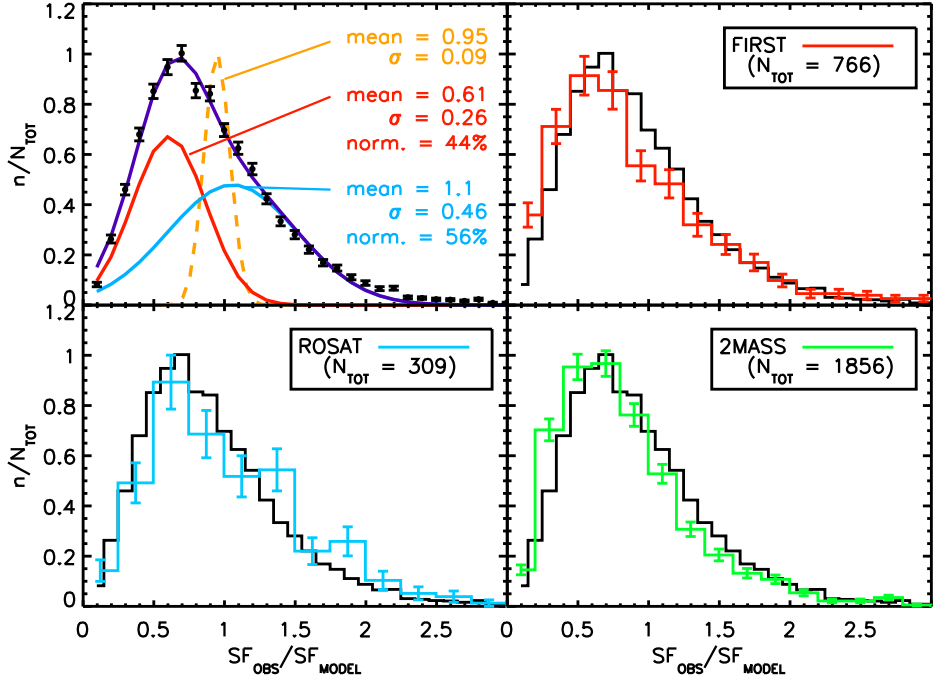}
\caption{Distribution of $SF_{obs}/SF_{model}$. The distribution from analysis by 
  I04 based on two-epoch data is shown by the dashed curve (scaled by a factor of 0.2). The symbols with error bars show the 
  distribution obtained here, that is based on SF measured for each 
  individual object ($n$ is defined as the number of points in a bin 
  ($N_{bin}$) divided by the bin width ($\Delta_{bin}$), and $N_{TOT}$ 
  is the total number of points used for each histogram; that is, the 
  enclosed area is unity). Error bars are computed as 
  $\sqrt{N_{bin}}/(N_{TOT}\Delta_{bin})$. The curve passing through
  the data points is an empirical fit that is the sum of the two solid 
  (Gaussian) curves below it, with relative normalizations of 44\% and
  56\%. The same total distribution is shown as 
  the dark histogram in the other three panels, where it is compared to
  analogous distributions for radio (\emph{top-right}),
  X-ray (\emph{lower-left}), and infrared (\emph{lower-right}) subsamples.}
\label{fig:hist}
\end{figure}

The distribution of $SF_{obs}/SF_{model}$ values for the two-epoch sample
analyzed by I04 is shown in the upper left panel in Fig.~\ref{fig:hist}. 
It is well described by a Gaussian distribution with $\sigma=0.09$. 
That is, the model describes the {\it mean} behavior of SF as a function
of luminosity, rest-frame wavelength and time to within $\sim$10\%
of the measured values. However, the analysis based on two-epoch data cannot provide information
about SF variance in a bin with fixed values of $M_i$, $\Delta t_{RF}$, and $\lambda_{RF}$. This is
because the SF is constructed using the magnitude difference measurements
from all the quasars. Their scatter measures the mean value of the SF, but
provides no information about the SF variance among individual objects.
To measure the latter, individual light curves must be available.

The distribution of $SF_{obs}/SF_{model}$ values for the sample
analyzed here, where SF is evaluated for every individual object
from its light curve, is also shown in the upper left panel in 
Fig.~\ref{fig:hist}. It is much wider than the corresponding 
distribution based on two-epoch sample (which covers a larger range in $\Delta t_{RF}$), clearly non-Gaussian, 
and well-fit by a sum of two Gaussians. In fact, the scatter in $SF_{obs}/SF_{model}$ for fixed values of $M_i$, $\Delta t_{RF}$, and $\lambda_{RF}$, is 
similar to the scatter induced by the variance of these quantities.  The published analysis, 
exemplified by VB, dVBWL and I04 work, captured the mean trends, but 
not the surprisingly large scatter around the mean behavior 
($\sigma\sim0.4$).  This new result demonstrates
that there is significant additional information that can be
extracted from well-sampled light curves for individual objects.

At this stage of analysis it is not clear what causes the SF
scatter. One obvious candidate is intrinsic stochasticity of 
the process causing variability (e.g. see discussion in 
\cite{Kawa1998}). We are also revisiting issues such as 
photometric calibration of SDSS stripe 82 data and the 
robustness of SF analysis to non-Gaussian outliers (the 
distribution of two-epoch magnitude differences follows 
an exponential, rather than Gaussian, distribution; I04).  It will also be helpful in the future to repeat this analysis for better-sampled light curves.

\subsection{ The SF for Subsamples with Detections at Other Wavelengths }

We investigated whether the distribution of $SF_{obs}/SF_{model}$
for the sample analyzed here varies among various subsamples. First
we compared the distribution for the whole sample to the distribution
obtained for the apparently brightest 10\%  objects in the $i$ 
band, and did not detect any statistically significant difference. We also created 
subsamples that are detected in 2MASS, FIRST, and ROSAT, as listed in
\cite{Sch2007}. Out of the 10,370 total data points, 1856 (or $\sim$400 
quasars) have infrared ($J$, $H$, and $K$) detections, 766 ($\sim$150 quasars) 
have radio detections, and 309 ($\sim$70 quasars) have X-ray detections.

\begin{figure}
\centering
\includegraphics[width=4.6in,viewport=38 315 575 730,clip]{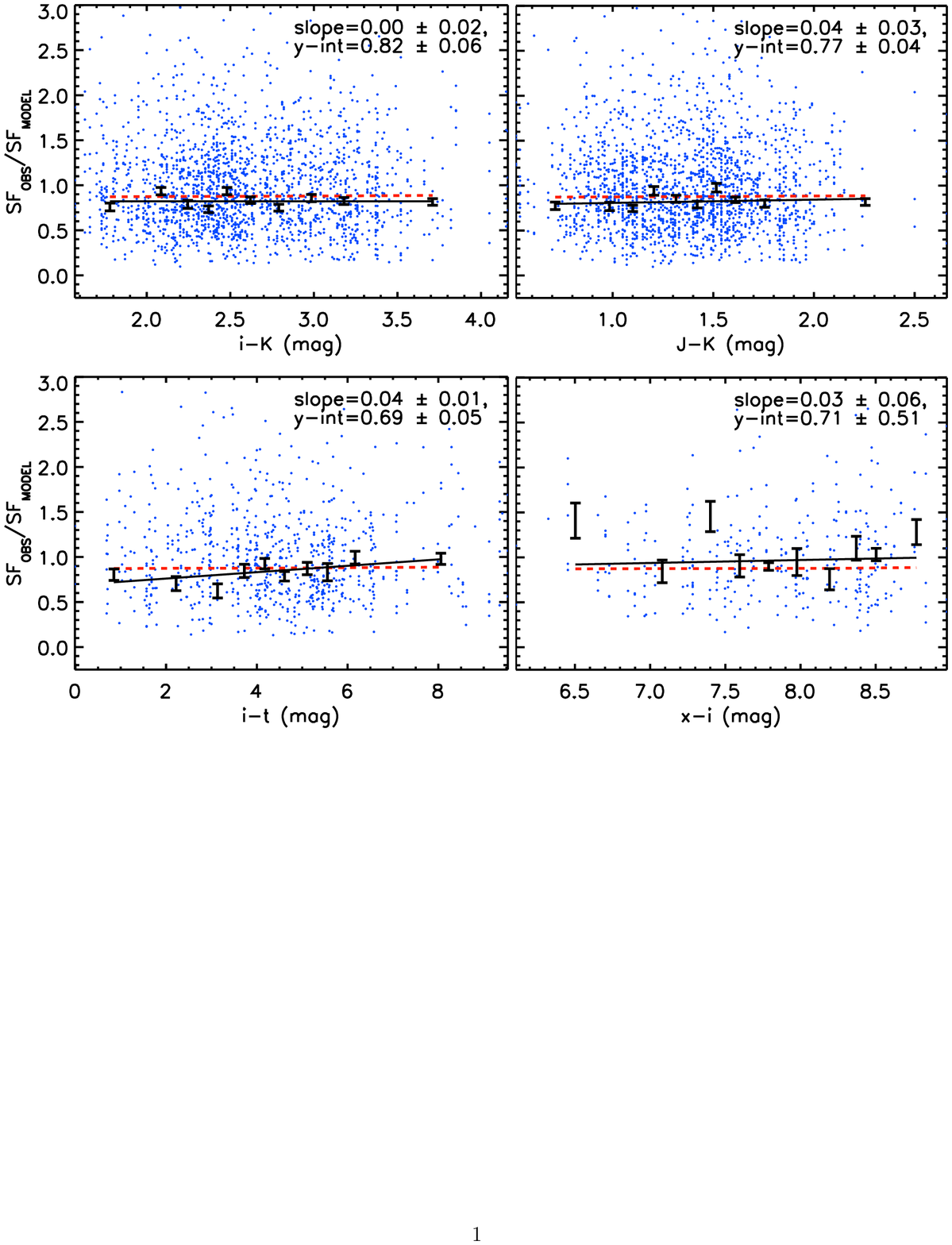}
\caption{The ratio of observed SF to model SF is plotted against $i-K$, $J-K$, $i-t$, and $x-i$ colors as dots.  These data points are then divided into bins of $N$ points, where $N$ is set to 10\% of the total number of points. The median values for each bin along with $1\sigma$ error bars are
  shown, where $\sigma = 0.926$(interquartile
  range)$/\sqrt{N-1}$. The over-plotted solid lines are linear
  regressions for the median values (the fit parameters are listed in
  the upper-right corners) and the over-plotted dashed lines are the
  same, but for the entire sample of 10,370 data points.}
\label{fig:scatter}
\end{figure}

The distribution of $SF_{obs}/SF_{model}$ for each of these
subsamples is compared to that of the entire sample in
Fig.~\ref{fig:hist}. While the ROSAT sample seems to
follow the distribution for the whole sample, the FIRST and 2MASS samples appear
skewed toward lower values of $SF_{obs}/SF_{model}$, indicating less variability
compared to the optically selected sample. We also investigated the behavior
of $SF_{obs}/SF_{model}$ values as a function of optical and infrared colors $i-K$ 
and $J-K$, optical-radio ``color'' $i-t$, and X-ray-optical ``color'' $x-i$ 
($t$ and $x$ are radio and X-ray AB magnitudes, see \cite{ive2002}). 
Figure~\ref{fig:scatter} shows that the quantity $SF_{obs}/SF_{model}$
is independent of $i-K$ and $J-K$ colors.  However, there seems to be
a positive correlation with $i-t$: optical variability increases with
radio loudness, in agreement with VB, who found that
radio-bright quasars are about 1.3 times more variable. There is no
significant correlation with $x-i$ color.

\end{document}